\documentclass[aps,prl,twocolumn,epsfig,groupedaddress]{revtex4}

\usepackage{epsfig}
\begin{document}

\title{Medium Modification of Jet Shapes and Jet Multiplicities}

\author{Carlos A. Salgado and Urs Achim Wiedemann}
\address{Theory Division, CERN, CH-1211 Geneva 23, Switzerland}

\date{\today}

\begin{abstract}
Medium-induced parton energy loss is widely considered to 
underly the suppression of high-$p_t$ leading hadron 
spectra in $\sqrt{s_{\rm NN}} = 200$ GeV Au+Au collisions 
at RHIC. Its 
description implies a characteristic 
$k_t$-broadening of the subleading hadronic fragments 
associated to the hard parton. However,
this latter effect is more difficult to measure and remained 
elusive so far. Here, we discuss how it affects genuine jet 
observables which are accessible at LHC and possibly at RHIC. 
We find that the $k_t$-broadening of 
jet multiplicity distributions provides a very sensitive probe of 
the properties of dense QCD matter, whereas the sensitivity
of jet energy distributions is much weaker. In particular, the
sensitive kinematic range of jet multiplicity distributions
is almost unaffected by the high multiplicity background. 
\end{abstract}
\maketitle
 \vskip 0.3cm


Hard partons produced in dense QCD matter are expected to loose
a significant fraction of their energy due to medium-induced
gluon radiation prior to hadronization~\cite{Gyulassy:1993hr}. 
This follows from calculations of the underlying non-Abelian 
Landau-Pomeranchuk-Migdal effect and allows to predict the 
dependence of parton energy loss on pathlength and density in 
a static~\cite{Baier:1996sk,Zakharov:1997uu,Wiedemann:2000za,Gyulassy:2000er}
or expanding~\cite{Baier:1998yf,Gyulassy:2000gk,Salgado:2002cd} medium.
Recent measurements~\cite{rhic}
of high-$p_t$ hadroproduction and its
centrality dependence in Au-Au collisions at $\sqrt{s_{\rm NN}} = 200$ GeV 
provide the first evidence~\cite{Wang:2003aw} for the occurrence of 
this jet quenching phenomenon. They allow to access properties of 
the dense medium produced in nucleus-nucleus collisions by analyzing 
the medium modification of high-$p_t$ 
hadroproduction~\cite{Gyulassy:2001nm,Wang:2002ri,Salgado:2002cd}.

So far, these analyzes are limited to the study of leading
hadron spectra and leading hadron back-to-back correlations.
However, energy loss of the leading parton implies
a redistribution of the associated
jet energy in transverse phase space or multiplicity. 
Thus, the observed energy degradation of leading hadrons should 
be reflected in the modification of genuine jet observables 
such as jet shapes and jet multiplicity distributions. 
The main aim of this letter is to calculate for the first time
medium-modified jet observables in the same theoretical framework
on which the current jet quenching interpretation of suppressed 
high-$p_t$ hadroproduction is based. 

\begin{figure}[h]\epsfxsize=8.7cm
\centerline{\epsfbox{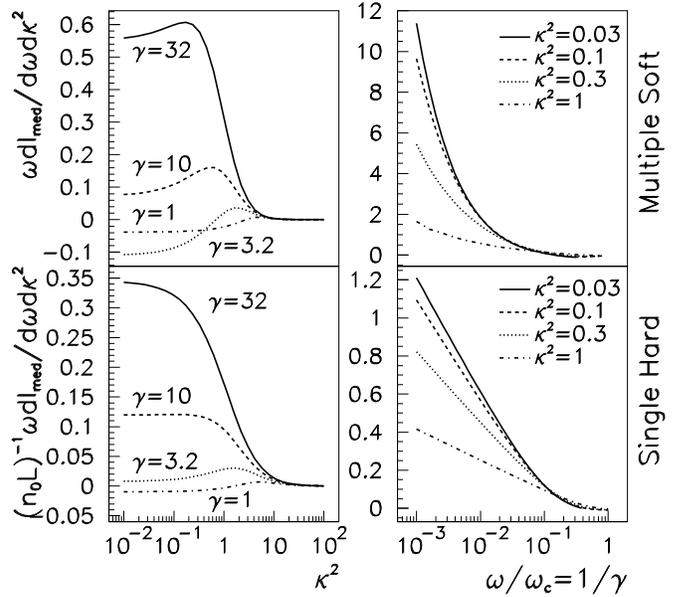}}
\caption{The gluon energy distribution 
(\protect\ref{eq1}) as a function of the rescaled gluon energy 
$\omega/\omega_c$ and the rescaled gluon transverse momentum 
$\kappa$.
}\label{fig1}
\end{figure}

We start
from the $k_t$-differential medium-induced 
distribution of gluons of energy $\omega$ radiated off an initial hard 
parton~\cite{Wiedemann:2000za,Wiedemann:2000tf,Salgado:2003gb},
\begin{eqnarray}
  &&\omega\frac{dI_{\rm med}}{d\omega\, d{\bf k}}
  = {\alpha_s\,  C_F\over (2\pi)^2\, \omega^2}\,
    2{\rm Re} \int_{0}^{\infty}\hspace{-0.3cm} dy_l
  \int_{y_l}^{\infty} \hspace{-0.3cm} d\bar{y}_l\,
   \int d^2{\bf u}\,  
  \nonumber \\
  &&\quad  \times
  e^{-i{\bf k}_t\cdot{\bf u}}   \,
  e^{ -\frac{1}{2} \int_{\bar{y}_l}^{\infty} d\xi\, n(\xi)\, \sigma({\bf u})}\,
  {\partial \over \partial {\bf y}}\cdot
  {\partial \over \partial {\bf u}}\,
  \nonumber \\
  && \quad \times \int_{{\bf y}={\bf r}(y_l)}^{{\bf u}={\bf r}(\bar{y}_l)} 
  \hspace{-0.5cm} {\cal D}{\bf r}
   \exp\left[ i \int_{y_l}^{\bar{y}_l} \hspace{-0.2cm} d\xi
        \frac{\omega}{2} \left(\dot{\bf r}^2
          - \frac{n(\xi)\, \sigma({\bf r})}{i\,2\, \omega} \right)
                      \right]\, .
    \label{eq1}
\end{eqnarray}
Medium properties enter (\ref{eq1}) via the product of the 
medium density $n(\xi)$ of scattering centers times the dipole 
cross section $\sigma({\bf r})$ which measures the 
interaction strength of a single elastic scattering. 
We first establish that Eq.~(\ref{eq1}) implies a 
one-to-one correspondence between the average energy loss
of the parent parton, and the transverse momentum broadening
of the associated gluon radiation, as argued in 
Ref.~\cite{Baier:1996sk}. To this end, we evaluate
$\omega\frac{dI_{\rm med}}{d\omega\, d{\bf k}}$ for 
$\alpha_s\, C_F = \frac{4}{9}$
in two approximations:

In the multiple soft scattering limit $n(\xi)\, \sigma({\bf r}) 
\approx \frac{1}{2} \hat{q}(\xi)\, {\bf r}^2$, the transport 
coefficient $\hat{q}$ characterizes the average transverse momentum 
squared transferred from the medium to 
the projectile per unit pathlength. In this case, medium-induced 
gluon radiation is limited to gluon energies 
$\omega < \omega_c = \frac{1}{2}\, \hat{q}\, L^2$,
see Fig.~\ref{fig1}. 
In medium pathlength $L$ and transport coefficient $\hat{q}$ 
determine not only the average energy
loss of the leading parton, $\Delta E = \int d\omega\, \omega
\frac{dI_{\rm med}}{d\omega} \sim \alpha_s \omega_c$, but also the 
typical transverse momentum transferred from the medium.
This limits medium-induced gluon radiation to 
$\kappa^2 = \frac{{\bf k}^2}{\hat{q}\, L} < 1$.    

%
%
The same conclusion is reached in the $N=1$ opacity expansion of
(\ref{eq1}) in which the medium is characterized by the average 
transverse momentum $\mu$ per scattering times the average number
$n_0\, L$ of such scatterings. In this case, the radiation is
limited to a characteristic gluon energy 
$\omega_c = \frac{1}{2} \mu^2\, L$ and a typical
transverse momentum $\kappa^2 = \frac{{\bf k}^2}{\mu^2}$.
The opacity $n_0\, L$ can be adjusted such that both approximations
give quantitatively comparable results for phase-space averaged
quantities as e.g. the average energy loss~\cite{Salgado:2003gb}. 
Differences in the
shape of the distributions shown in Fig.~\ref{fig1} are indicative
of the uncertainties in modeling the detailed structure of the medium.

\begin{figure}[h]\epsfxsize=8.5cm
\centerline{\epsfbox{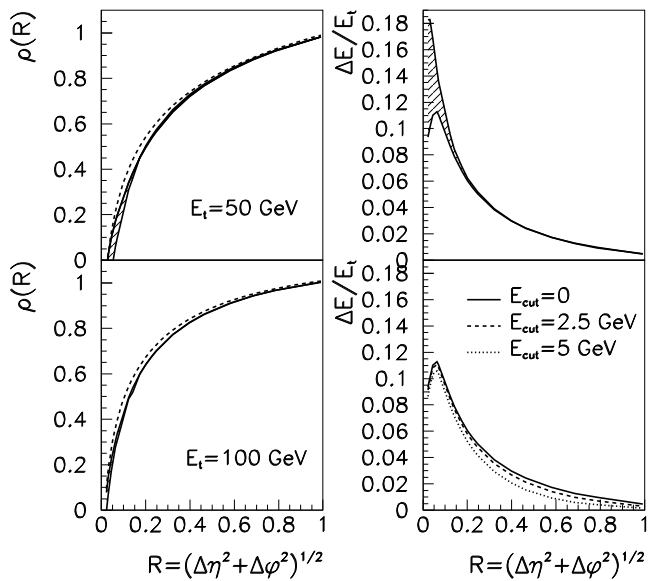}}
\caption{LHS: The jet shape (\protect\ref{eq5}) for a 50 GeV and
100 GeV quark-lead jet which fragments in the vacuum (dashed curve) or 
in a dense QCD medium (solid curve)
characterized by $\omega_c = 62$ GeV and $\omega_c\, L = 2000$.
RHS: the corresponding
average medium-induced energy loss for $E_t = 100$ GeV
outside a jet cone $R$ radiated 
away by gluons of energy larger than $E_{\rm cut}$. Shaded regions 
indicate theoretical uncertainties discussed in the text.
}\label{fig2}
\end{figure}

Next we ask to what extent the $k_t$-broadening of the
medium-modified parton shower, established in Fig.~\ref{fig1}, 
shows up in the azimuthal redistribution of jet energy.
We start form the fraction $\rho(R)$ of the total jet energy 
$E_t$ deposited within a given jet subcone of radius 
$R = \sqrt{(\Delta \eta)^2 + (\Delta \Phi)^2}$,
\begin{eqnarray}
  \rho_{\rm vac}(R) &=& \frac{1}{N_{\rm jets}} \sum_{\rm jets}
  \frac{E_t(R)}{E_t(R=1)}\, .
  \label{eq5}
\end{eqnarray}
In the absence of medium-effects, this jet shape is described e.g. 
by the parametrization~\cite{Abbott:1997fc} of the Fermilab 
$D0$ Collaboration for jets in the range $\approx 50 < E_t < 150$ 
GeV and opening cones $0.1 < R < 1.0$. In what follows, we work in 
the dijet center of mass where the jet width in pseudorapidity 
$\Delta \eta$ and azimuth $\Delta \Phi$ is related to the gluon 
emission angle $\Theta$ of our calculation as $R = \Theta$. 
To discuss the medium-dependence of $\rho(R)$, we
calculate the probability $P_{\rm tot}(\epsilon,\Theta)$ that 
a fraction $\epsilon$ of the total jet energy $E_t$ is
emitted outside the angle $\Theta$. Assuming that
gluon emission follows an independent Poisson process, this
probability is given by~\cite{Baier:2001yt}
\begin{eqnarray}
  &&\hspace*{-0.4cm}
  P_{\rm tot}(\epsilon,\Theta) 
  = \int_C \frac{d\nu}{2\pi i}\, e^{\nu\, \epsilon}\, 
%
  \nonumber \\
  &&\hspace*{-0.4cm} \times 
  \exp\left[ -\int_0^\infty d\omega\, 
    \left( \frac{dI^{>\Theta}_{\rm vac}}{d\omega} 
            + \frac{dI^{>\Theta}_{\rm med}}{d\omega} \right)
    \left( 1-e^{-\nu\omega}\right)\right]\, ,
  \label{eq2}
\end{eqnarray}
where the contour $C$ goes along the imaginary axis. The expression 
(\ref{eq2}) takes into account the angular energy distribution 
of the parton fragmentation process in the vacuum, 
$\frac{dI^{>\Theta}_{\rm vac}}{d\omega} = \int_\Theta^\pi\, d\varphi\, 
\frac{dI_{\rm vac}}{d\omega\, d\varphi}$, as well as its
medium-modification $\frac{dI^{>\Theta}_{\rm med}}{d\omega}$
calculated from eq. (\ref{eq1}). 
Since both contributions are additive, the total probability 
(\ref{eq2}) can be written as a convolution of the vacuum
and the medium-induced probability
\begin{equation}
  P_{\rm tot}(\epsilon,\Theta) = \int d\epsilon_1\, 
  P_{\rm vac}(\epsilon_1,\Theta)\,  
  P_{\rm med}(\epsilon - \epsilon_1,\Theta)\, .
  \label{eq3}
\end{equation} 
We calculate the quenching weight $P_{\rm med}(\epsilon,\Theta)$ from
eq. (\ref{eq1}), see Ref.~\cite{Salgado:2003gb}.
The vacuum contribution $P_{\rm vac}(\epsilon,\Theta)$ in (\ref{eq3}) 
is determined by the experimentally measured jet shape 
$\rho_{\rm vac}(R)$, normalized to the vacuum fraction of the
total jet energy
\begin{equation}
  \int d\epsilon\, \epsilon\, P_{\rm vac}(\epsilon,\Theta)
  = \frac{E_t-\Delta E}{E_t} \left[ 
  1-\rho_{\rm vac}(R=\Theta) \right]\, ,
  \label{eq6}
\end{equation}
where $\Delta E \equiv \Delta E(R=0) = \int d\epsilon \epsilon
P_{\rm med}(\epsilon,\Theta = 0)$.
In the absence of tabulated experimental data on the width of
$P_{\rm vac}(\epsilon,\Theta)$, we choose a sharply peaked
distribution $P_{\rm vac}(\epsilon,\Theta) =
\delta\left(\epsilon -\frac{E_t-\Delta E}{E_t}
\left[1-\rho_{\rm vac}(R)\right]\right) 
\vert_{R=\Theta}$. The medium-modified jet shape 
$\rho_{\rm med}(R)$ 
is then defined in terms of the average jet energy fraction 
$\frac{\Delta E(\Theta)}{E_t}$ radiated outside an angle $\Theta$,
\begin{eqnarray}
  &&\rho_{\rm med}(R) \equiv 
  1 - \int d\epsilon\, \epsilon\, P_{\rm tot}(\epsilon,\Theta=R)
  \nonumber \\
  && \quad =  \rho_{\rm vac}(R) - \frac{\Delta E_t(R)}{E_t}
        + \frac{\Delta E}{E_t} \left( 1 - \rho_{\rm vac}(R)\right)\, ,
      \label{eq9}
\end{eqnarray}
We have calculated (\ref{eq9}) as a function of the in-medium 
pathlength $L$, the jet energy $E_t$, 
and the transport coefficient $\hat{q}$. Numerical results are shown in
Fig.~\ref{fig2}. In the eikonal approximation, the quenching weight 
$P_{\rm med}(\epsilon,\Theta)$ is known to
have support in the unphysical region $\epsilon > 1$~\cite{Salgado:2003gb}. 
This introduces an uncertainty which we estimate with the shaded region
in Fig.~\ref{fig2} by comparing
the result of an unrestricted $\epsilon$
integration in (\ref{eq9}), $\frac{\Delta E(\Theta)}{E_t}\vert_1 \equiv
\int d\epsilon\, \epsilon\, P_{\rm med}(\epsilon,\Theta)$, to 
the properly reweighted restricted integration
\begin{equation}
  \frac{\Delta E(\Theta)}{E_t}\Bigg\vert_2 = 
  \frac{\int_0^1 d\epsilon\, \epsilon\, P_{\rm med}(\epsilon,\Theta)}
  {\int_0^1 d\epsilon\, P_{\rm med}(\epsilon,\Theta)}\, .
\end{equation}
In general, we find that the medium modification (\ref{eq9}) 
grows approximately linear with the transport coefficient (data not shown)
in agreement with the $\hat{q}$-dependence of the average energy 
loss $\Delta E(\Theta)$. The 
medium modification decreases approximately like $1/E_t$ with 
increasing jet energy. Qualitatively comparable results are 
obtained  in the $N=1$ opacity approximation (data not shown).

%

The parameter values chosen in Fig.~\ref{fig2} ($\omega_c = 62$ GeV, 
$\omega_c\, L= 2000 $) amount to an average squared momentum transfer 
of $\hat{q}\, L \approx (2\, {\rm GeV})^2$  from the medium to the partonic 
jet components. This is a rough estimate for nucleus-nucleus 
collisions at the LHC and corresponds to an initial gluon density which
is a factor $\approx 2$ larger than values extracted from
RHIC data~\cite{Salgado:2002cd}. The resulting medium-induced 
broadening of the jet shape
shown in  Fig.~\ref{fig2}, amounts to a
slightly reduced average jet energy fraction inside small jet 
cones $R = 0.3$ by $\sim$ 5 \%
for $E_t = 50$ GeV and $\sim$ 3 \% for $E_t = 100$ GeV.
Although somewhat larger initial gluon densities and thus larger 
medium modifications are conceivable at LHC, the size of the 
resulting modification remains small. 
For larger jet cones ($R > 0.7$ say),
medium effects become even smaller since the
medium-induced energy redistribution occurs mainly inside 
the jet cone. This may allow to measure the total jet energy 
above background without resorting to jet samples ``tagged'' 
by a recoiling hard photon or $Z$-boson. It implies that jet 
$E_t$ cross sections in nucleus-nucleus collisions scale 
with the number of binary collisions.

In nucleus-nucleus collisions at LHC, jets up to $E_t > 200$  GeV 
will be produced abundantly~\cite{Accardi:2002vt}. However, 
the background $E_t^{\rm bg}$ deposited 
inside the corresponding jet cone (for event multiplicity 
$dN^{\rm ch}/dy = 2500$, we estimate $E_t^{\rm bg} \sim 100$ GeV
for $R = 0.3$ and $E_t^{\rm bg} \sim 250$ GeV for $R=0.5$)
is of comparable size. Thus, it is feasible to disentangle a
high-$E_t$ jet from background, 
but the measurement of a $< 10$ \% modification of its
shape remains challenging. In particular, such precision
may require a better theoretical understanding of how the initial
state radiation associated to a high-$E_t$ jet affects the
underlying event and its fluctuations 
(the so-called pedestal effect). 

\begin{figure}[t!]\epsfxsize=8.7cm
\centerline{\epsfbox{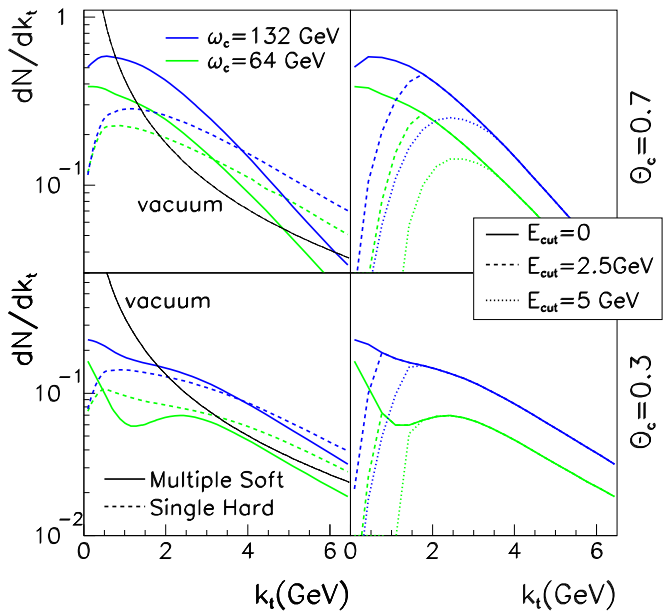}}
\caption{Comparison of the vacuum and medium-induced part of the 
gluon multiplicity distribution
(\protect\ref{eq8}) inside a cone size $R=\Theta_c$, measured as a 
function of $k_t$ with respect to the jet axis. Removing gluons with
energy smaller than $E_{\rm cut}$ from the distribution (dashed and dotted
lines) does not affect the high-$k_t$ tails.
}\label{fig3}
\end{figure}

Interestingly, the transverse momentum broadening shown
in Fig.~\ref{fig2} changes only weakly with a low momentum
cut-off which removes gluon emission below 5 GeV. This can be
understood in terms of formation time and phase space limitations
in a small-size medium~\cite{Salgado:2003gb}.
As a consequence, the transverse momentum broadening of $\rho(R)$ 
is mainly due to high energy partons which can be expected to
contribute significantly to the hadron yield above background. 
To study this point in more detail, we
have calculated the medium-induced additional number of gluons 
with transverse momentum $k_t = \vert {\bf k}\vert$, produced 
within a subcone of opening angle $\theta_c$, 
\begin{eqnarray}
 \frac{dN_{\rm med}}{dk_t} =  \int_{k_t/\sin\theta_c}^{E_t} d\omega\,
               \frac{dI_{\rm med}}{d\omega\, dk_t}\, .
  \label{eq8}
\end{eqnarray}
In Fig.~\ref{fig3}, we compare this distribution to the shape
of the corresponding vacuum component, 
$\frac{dN_{\rm vac}}{dk_t} \propto \frac{1}{k_t}\, 
\log(E_t\sin\theta_c/k_t)$, calculated from eq. (\ref{eq1}) as well.
The total partonic jet multiplicity is the sum of both components.
For realistic values of medium density and in-medium pathlength,
medium effects are seen to increase this multiplicity significantly 
(by a factor $\sim 2 - 5$) in particular in the high-$k_t$ tails. 
Also, the shape and width of the distribution (\ref{eq8}) changes
sensitively with the scattering properties of the medium.
Moreover, since gluons must have a minimal energy $\omega > 
k_t/\sin\Theta_c$ to be emitted inside the jet cone, this 
high-$k_t$ tail is unaffected by ``background'' cuts on the 
soft part of the spectrum, see Fig.~\ref{fig3}. These qualitative
conclusions are not affected by the uncertainties of our calculation
which are illustrated by the significant differences in    
the angular dependence of the medium-induced gluon
radiation (\ref{eq1}) in the multiple soft and single hard
scattering approximation~\cite{Salgado:2003gb}. In particular,
destructive interference effects are known~\cite{Wiedemann:2000tf} 
to be more significant
in the multiple soft scattering limit and for small angles, which
may explain the non-monotonous behaviour seen for $\Theta_c = 0.3$
in Fig.~\ref{fig3}. 

%

On the basis of Fig.~\ref{fig3}, we argue that the measurement
of the transverse momentum distribution of hadrons with respect
to the jet axis is very sensitive to the transverse momentum
broadening of the underlying parton shower and should be detectable
above background. Despite the expected enhancement in the high-$k_t$
tail, the total multiplicity of a jet will increase and the average
energy of the hadronic fragments will soften.
Hadronization of the parton shower is known to
affect the absolute size and shape of this multiplicity distribution in the
vacuum~\cite{Acosta:2002gg} and can be expected to modify the 
medium-dependent part as well. Also, experimental effects such as an
increased uncertainty in determining the jet axis in a high multiplicity
environment tend to broaden the distribution and have to be taken into
account properly. To become more quantitative on the level of hadronic
observables requires presumably a
Monte Carlo implementation of the medium-modified parton shower 
which is not at our disposal yet. However, the 
effect observed in Fig.~\ref{fig3} can be expected to survive
hadronization. In particular, the insensitivity of
the high-$k_t$ tail to the low $E_t$ background and its
sensitivity to the transverse momentum picked up from the medium
are both based on kinematic grounds and should not depend 
on the details of our calculation.  

Other multiplicity distributions may show interesting medium 
modifications as well. As an example, we mention the
modifications of the hump-backed rapidity plateau,
i.e. the number of hadrons with jet energy fraction $x$ inside
the jet. In the vacuum, it is well-described by the result of 
the MLLA approximation which depends on the jet energy only via the
parameter combination $\frac{E_t\, \sin\Theta_c}{Q_{\rm eff}}$
with $Q_{\rm eff} \sim 250\, {\rm MeV} \sim \Lambda_{\rm QCD}$ the 
only fit parameter~\cite{Acosta:2002gg}.
The medium-modification of the corresponding partonic quantity is 
\begin{eqnarray}
 \frac{dN_{\rm med}}{d\log x} 
  =  \int_0^{\frac{x\,E_t\, \sin\Theta_c}{\sqrt{\hat{q}L}}} d\kappa^2\,
               \frac{dI_{\rm med}}{d\log x\, d\kappa^2}\, .
  \label{eq15}
\end{eqnarray}
Here, we only observe that the medium modification of 
$\frac{dN_{\rm med}}{d\log x}$ supplements the non-perturbative
scale $ Q_{\rm eff}$ with a perturbatively large scale 
$\sqrt{\hat{q}L} \sim Q_s$. A more detailed analysis of (\ref{eq15}) 
and other multiplicity distributions is left to future work.

We finally comment on the implications of our study for the
ongoing experiments at RHIC. In general, the strategy of  
triggering on the most energetic hadron biases
jet samples significantly and may deplete in particular the
multiplicity of subleading high-momentum hadrons. Furthermore,
if the energy of the leading particle is not sufficiently high,
the transverse phase space mapped out in Fig.~\ref{fig3} is
simply not available. However, our study points to the possibility
that a significant increase in jet multiplicity (and hence a 
decrease of the average energy of the leading hadron inside the jet) is
accompanied by a rather moderate change in the angular distribution
of the jet energy flow. This may be tested at RHIC e.g. by measuring
in back-to-back dihadron correlations the total $E_t$ 
(or multiplicity) in a cone around the triggered hadron as well as 
the balancing energy in the
opposite direction, and subtracting the background energy
$E_t^{\rm bg}$. 

In summary, 
while $k_t$-broadening in the initial~\cite{Accardi:2003jh} 
and final~\cite{Qiu:2003pm} state has been discussed repeatedly 
for leading hadron spectra, the observables studied
here relate this broadening quantitatively to parton energy loss.
Their measurement would not only further substantiate the picture 
of a medium-modified parton shower which underlies the current jet 
quenching interpretation of high-$p_t$ hadroproduction.
Compared to leading hadron spectra,
the $k_t$-broadening of multiplicity distributions may
also provide data of competing accuracy for a better tomographic
characterization of dense QCD matter.

We thank Nestor Armesto, Rolf Baier, Andreas Morsch, J\"urgen Schukraft,
Fuqiang Wang and Bolek Wyslouch for helpful discussion. In particular,
we thank J\"urgen Schukraft for pointing out an error in an earlier
version of this work.

%


\begin{thebibliography}{9}
%
\bibitem{Gyulassy:1993hr}
M.~Gyulassy and X.~N.~Wang,
Nucl.\ Phys.\ B {\bf 420} (1994) 583.
%
\bibitem{Baier:1996sk}
R.~Baier, Y.~L.~Dokshitzer, A.~H.~Mueller, S.~Peigne and D.~Schiff,
Nucl.\ Phys.\ B {\bf 484} (1997) 265.
%
\bibitem{Zakharov:1997uu}
B.~G.~Zakharov,
JETP Lett.\  {\bf 65} (1997) 615.
%
\bibitem{Wiedemann:2000za}
U.~A.~Wiedemann,
Nucl.\ Phys.\ B {\bf 588} (2000) 303.
%
\bibitem{Gyulassy:2000er}
M.~Gyulassy, P.~Levai and I.~Vitev,
Nucl.\ Phys.\ B {\bf 594} (2001) 371.
%
\bibitem{Baier:1998yf}
R.~Baier, Y.~L.~Dokshitzer, A.~H.~Mueller and D.~Schiff,
Phys.\ Rev.\ C {\bf 58} (1998) 1706.
%
\bibitem{Gyulassy:2000gk}
M.~Gyulassy, I.~Vitev and X.~N.~Wang,
Phys.\ Rev.\ Lett.\  {\bf 86} (2001) 2537.
%
\bibitem{Salgado:2002cd}
C.~A.~Salgado and U.~A.~Wiedemann,
Phys.\ Rev.\ Lett.\  {\bf 89} (2002) 092303
%
\bibitem{rhic}
K.~Adcox {\it et al.}  [PHENIX Coll.],
Phys.\ Rev.\ Lett.\  {\bf 88} (2002) 022301;
%
S.~S.~Adler  [PHENIX Coll.],
arXiv:nucl-ex/0308006;
%
C.~Adler {\it et al.},
Phys.\ Rev.\ Lett.\  {\bf 89} (2002) 202301;
%
J.~Adams {\it et al.},  [STAR Coll.],
arXiv:nucl-ex/0305015.
%
\bibitem{Wang:2003aw}
X.~N.~Wang,
arXiv:nucl-th/0307036.
%
\bibitem{Gyulassy:2001nm}
M.~Gyulassy, P.~Levai and I.~Vitev,
Phys.\ Lett.\ B {\bf 538} (2002) 282.
%
\bibitem{Wang:2002ri}
E.~Wang and X.~N.~Wang,
Phys.\ Rev.\ Lett.\  {\bf 89} (2002) 162301.
%
\bibitem{Wiedemann:2000tf}
U.~A.~Wiedemann,
Nucl.\ Phys.\ A {\bf 690} (2001) 731.
%
\bibitem{Salgado:2003gb}
C.~A.~Salgado and U.~A.~Wiedemann,
Phys.\ Rev.\ D {\bf 68} (2003) 014008.
%
\bibitem{Abbott:1997fc}
B.~Abbott, M.~Bhattacharjee, D.~Elvira, F.~Nang and H.~Weerts  [D0 Coll.],
FERMILAB-PUB-97-242-E
%
\bibitem{Baier:2001yt}
R.~Baier, Y.~L.~Dokshitzer, A.~H.~Mueller and D.~Schiff,
JHEP {\bf 0109} (2001) 033.
%
\bibitem{Accardi:2002vt}
A.~Accardi, N.~Armesto and I.~P.~Lokhtin,
arXiv:hep-ph/0211314.

\bibitem{Acosta:2002gg}
D.~Acosta {\it et al.}  [CDF Coll.],
FERMILAB-PUB-02-096-E
%
\bibitem{Accardi:2003jh}
A.~Accardi and M.~Gyulassy,
arXiv:nucl-th/0308029.
%
\bibitem{Qiu:2003pm}
J.~W.~Qiu and I.~Vitev,
Phys.\ Lett.\ B {\bf 570} (2003) 161.


\end{thebibliography}
\end{document}